\pdfoutput=1
\documentclass[11pt]{article}
\usepackage{acl}
\usepackage{times} 
\usepackage{latexsym}
\usepackage[T1]{fontenc}
\usepackage[utf8]{inputenc}
\usepackage{microtype}
\usepackage{inconsolata}
\usepackage{graphicx}
\usepackage{amsmath,amsfonts}
\usepackage{algorithmic}
\usepackage{graphicx}
\usepackage{textcomp}
\usepackage{xcolor}
\usepackage{comment}
\usepackage{balance}
\usepackage{xspace}
\usepackage[ruled, lined, linesnumbered, commentsnumbered, longend]{algorithm2e}
\usepackage{multirow,multicol}
\usepackage{colortbl}
\usepackage{tabulary}
\usepackage[flushleft]{threeparttable}
\usepackage{caption}
\usepackage{subfigure}
\usepackage{tcolorbox}
\usepackage{mathtools}
\usepackage{url}
\usepackage{circledsteps}
\usepackage{tikz}
\usepackage{units}
\usepackage{diagbox}
\usepackage{listings}
\usepackage{enumitem}
\usepackage{ulem}
\usepackage{bbding}
\definecolor{backcolour}{rgb}{0.95,0.95,0.92}

\SetCommentSty{mycommfont}

\newcommand*\circled[1]{\tikz[baseline=(char.base)]{
            \node[shape=circle,fill,inner sep=1pt,font=\footnotesize] (char) {\textcolor{white}{#1}};}}

\usepackage{tabularx}
\usepackage{listings}
\usepackage{color}

\definecolor{dkgreen}{rgb}{0,0.6,0}
\definecolor{gray}{rgb}{0.5,0.5,0.5}
\definecolor{mauve}{rgb}{0.58,0,0.82}

\graphicspath{ {./figures/} }
\usepackage{textgreek}
\usepackage{float}

\newcommand{\taskone}{Assertion Generation\xspace}
\newcommand{\tasktwo}{Bugs2Fix\xspace}
\newcommand{\taskthree}{Code Suggestion\xspace}
\newcommand{\ourtool}{\textsc{CodePromptZip}\xspace}

\AtBeginDocument{%
  \providecommand\BibTeX{{%
    \normalfont B\kern-0.5em{\scshape i\kern-0.25em b}\kern-0.8em\TeX}}}

\begin{document}

\title{\ourtool: Code-specific Prompt Compression for Retrieval-Augmented Generation in Coding Tasks with LMs}


\author{Pengfei He \\
  University of Manitoba \\
  \texttt{hep2@myumanitoba.ca} \\\And
  Shaowei Wang \\
  University of Manitoba \\
  \texttt{shaowei.wang@umanitoba.ca} \\\And
  Tse-Hsun (Peter) Chen \\
  Concordia University \\
  \texttt{peterc@encs.concordia.ca} \\
  }




\maketitle

\begin{abstract}
Retrieval-Augmented Generation (RAG) enhances coding tasks by incorporating retrieved code examples into prompts. However, lengthy prompts—often exceeding tens of thousands of tokens—introduce challenges related to limited context windows of language models (LMs) and high computational costs. Existing prompt compression techniques focus on natural language, lacking tailored solutions for code. To address the gap, we propose \ourtool, a framework that compresses code examples before integrating into RAG workflows. Our framework employs a type-aware, priority-driven strategy to construct training samples for training code compression model. By using program analysis, we identify token types (e.g., Identifier) and perform ablation analysis to rank their removal priorities based on their impact on task performance. We then train a small LM as the compressor on these samples, enabling flexible compression conditioned on specified ratios while minimizing performance degradation. Specially, the compressor’s architecture is augmented with a copy mechanism, allowing tokens to be directly copied from the original code snippets. Evaluation results show that \ourtool surpasses SOTA entropy-based and distillation-based baselines, improving by 23.4\%, 28.7\%, and 8.7\% over the best baseline for \taskone, \tasktwo, and \taskthree, respectively. 


\end{abstract}

\section{Introduction}\label{sec:intro}

Retrieval-Augmented Generation (RAG) for language models \citep{rag,rag2,recomp} has shown remarkable performance on knowledge-intensive tasks, particularly in coding domains \citep{UBC,allyouneed,he2024}, by incorporating retrieved code examples into input prompts. However, such prompts often span tens of thousands of tokens, which creates challenges due to the limited context window of LMs and the high cost of processing long prompts with proprietary services like GPT-4 (\$2.50 per million tokens).

Prompt compression offers a promising solution for efficient LM utilization  by retaining essential information while reducing prompt length \citep{efficient}. Although existing studies have achieved promising results for natural language (NL) tasks, including language modeling \citep{recomp,soft,gist}, question-answering \citep{rl}, and summarization \citep{llmlingua,selective}, there is no compressor specifically for coding tasks. To address this gap, we introduce \ourtool, a framework to train a code-specific compressor to compress code examples for RAG-based coding tasks. 

We propose using a small LM (i.e., CodeT5 \citep{codet5}, 775M) as the compressor to compress code examples. The LM-based compressor captures the probabilistic relationships between code tokens without being constrained by strict syntax, making our framework applicable to incomplete code. The generated compressed examples aim to be lightweight yet effective, ensuring minimal impact on the base LM’s ability to produce high-quality outputs. To provide flexibility, the compressor accepts original code examples and desired compression ratios as input, generating examples that align with specified constraints. However, training the compressor introduces two key challenges: \circled{1} Constructing suitable datasets tailored for code compression. \circled{2} Designing a compressor architecture that effectively supports code compression while allowing compression ratio control.

To address \circled{1}, we propose a type-aware, priority-driven method to construct code compression datasets. This approach leverages the observation that different token types (e.g., Identifier) in code examples  have varying impacts on generation quality. Using program analysis tools \citep{codelm}, tokens are first categorized by their type. Next, we perform an ablation analysis to measure the impact of each type of tokens and establish a hierarchy of removal priorities based on their impact on performance degradation. Finally, a greedy strategy is employed to iteratively remove higher priority tokens and generate compressed code snippets with varying compression ratios.

To address \circled{2}, we enhance the base CodeT5 architecture with a copy mechanism \citep{copy,copy1}, which enables the model to directly copy tokens from the source. Since the compressed code is fully derived from the original, this mechanism introduces a copy distribution over source tokens to guide the token generation from the source sequence during decoding. Additionally, we extend the vocabulary by incorporating special tokens (e.g., <Ratio>), allowing the model to condition on specified compression ratios and adaptively learn compression at varying levels during training.

We evaluated \ourtool by compressing code examples in three RAG-based coding tasks, i.e., \taskone~\cite{UBC}, \tasktwo~\cite{codexglue}, and \taskthree~\cite{allyouneed}. \ourtool effectively maintains performance while reducing prompt lengths. \ourtool demonstrates improvements over both SOTA entropy-based (e.g., LLMLingua~\citep{llmlingua}) and distillation-based baselines (e.g., RECOMP~\citep{recomp}). \ourtool achieves an improvement of 23.4\%, 28.7\%, and 8.7\% over the best baseline for three coding tasks \taskone, \tasktwo, and \taskthree, respectively.



We make the following contributions.
\vspace{-0.1in}
\begin{itemize}
    \setlength\itemsep{-0.07in} 
    \item We first observe that different types of tokens have varying impacts on final generation quality. Based on this, we propose a novel prompt compression framework designed for compressing code examples.
    \item We developed a copy-enhanced LM as the compressor to compress code examples effectively and allow compression ratio control.
    \item Our approach achieves significant performance improvements over SOTA baselines and demonstrates generalization across different language models and tasks.
\end{itemize}

\section{Related Work and Background}
\subsection{Related work}\label{sec:related}


Prompt compression methods can be broadly classified into two types: \textbf{soft prompts} and \textbf{discrete prompt compression} \citep{efficient}. \textbf{Soft prompts} learn embeddings that encode either task instructions~\citep{gist} or example documents~\citep{soft}. For example, \citet{gist} condense prompt instructions into reusable ``gist'' vectors, while \citet{soft} compress long documents into learnable context vectors. However, soft prompts face limitations in cross-model compatibility and require gradient access to base LMs, making them impractical for API-based proprietary LM services.

Recent research has focused on \textbf{discrete prompt compression}, which retains key tokens from the original prompt while eliminating less informative content. This approach enhances compatibility with black-box or proprietary LMs. Notable techniques include \textbf{entropy-based} and \textbf{knowledge distillation} methods.
\textbf{Entropy-based} methods, such as LLMlingua and LongLLMlingua \citep{llmlingua,longllmlingua}, use small LMs to estimate the information entropy of tokens, filtering out low-value content. However, they rely on heuristic metrics that may not align well with compression objectives.
\textbf{Knowledge distillation} leverages large LMs like GPT-4 \citep{gpt4} to generate compressed summaries, which are then used to fine-tune smaller LMs as compressors. For instance, \citet{recomp} trained a T5 model on GPT-3.5-turbo summaries, while \citet{lingua2} employed a transformer encoder to classify tokens for extraction. Despite their effectiveness, distillation methods struggle with maintaining strict compression ratios and entail high costs due to reliance on proprietary LMs.

Although code is a subset of natural language, it exhibits unique features, such as type information \citep{codelm}. Different token types encapsulate distinct symbolic and syntactic information. For example, \textbf{Identifier} tokens reflect developers’ intent, while \textbf{Symbol} tokens define delimiters and operations. A recent work \citep{less} primarily targets the natural language parts (i.e., docstrings) in coding task prompts rather than addressing the compression of code itself. To the best of our knowledge, we are the first to focus on compressing the code.

\subsection{Problem Formulation}
Referring to \citealp{llmlingua}, we modify and reformulate prompt compression. For a coding task $\mathcal{T}$, given an original prompt, denoted as $\textbf{\textit{x}} = (\textbf{\textit{x}}_1^{code}, ..., \textbf{\textit{x}}_N^{code}, \textbf{\textit{x}}^{ques})$, where $\textbf{\textit{x}}_i^{code}$ represents $ith$ code example \footnote{As improving the retriever is not the focus of this work, we retrieve examples using the SOTA BM25 \citep{he2024}.}, $N$ represents number of shots, and $\textbf{\textit{x}}^{ques}$ represents the question. We aim to compress the code examples to reduce token count while retaining critical information for the question. Formally, the compression is performed by a compressor $\mathcal{LM_C}$, acting as a function:

\setlength{\abovedisplayskip}{7pt} 
\setlength{\belowdisplayskip}{5pt} 
\begin{align}
\widetilde{\textbf{\textit{x}}}_i^{code}=\mathcal{LM_C}(\mathcal{T},\tau_{code},\textbf{\textit{x}}_i^{code})
\end{align}\label{eq:lmc}

\noindent where $\tau_{code}=1-|\widetilde{\textbf{\textit{x}}}_i^{code}|/|\textbf{\textit{x}}_i^{code}|$ is the compression ratio for a code snippet, with $\mathcal{T}$ representing specific task contexts to account for variations in token importance across tasks.  vary cross tasks,  is . With compressed code examples, the overall prompt is shortened as:

\begin{align}
\widetilde{\textbf{\textit{x}}} &= \ourtool(\textbf{\textit{x}}) \\
&= (\{\mathcal{LM_C}(\mathcal{T},\tau_{code},\textbf{\textit{x}}_i^{code})\}_{i=1}^N, \textbf{\textit{x}}^{ques}) \nonumber
\end{align}

\noindent where the overall ratio is given by $\tau=1-\widetilde{\textbf{\textit{x}}}/\textbf{\textit{x}}$.
The generation of the base language model $\mathcal{\mathcal{BLM}}$ with the compressed prompt $\widetilde{\textbf{\textit{x}}}$ is expected to closely approximate the generation with the original prompt \textbf{\textit{x}}. This can be formulated as: 
\begin{align}
\min _{\widetilde{\boldsymbol{x}}, \tau} \operatorname{KL}\left(P\left(\mathcal{BLM}(\widetilde{\textbf{\textit{x}}})\mid \widetilde{\textbf{\textit{x}}}\right), P\left(\mathcal{BLM}(\textbf{\textit{x}}) \mid \textbf{\textit{x}}\right)\right)
\end{align}


\section{Type-aware Priority Ranking}\label{sec:ranking}

Before introducing our framework, we outline its motivation. Different tokens in a prompt contribute unevenly to the final output, with less impactful tokens prioritized for removal~\cite{lessismore,selective,llmlingua,longllmlingua,recomp}. In coding tasks, prompts often include code snippets as RAG demonstrations. A key question arises: \textit{Do different types of tokens in code contribute differently to the final results?} If so, this insight can guide code prompt compression. To explore this, we use program analysis (PA) tools to categorize tokens by type and conduct ablation analysis to identify those with minimal impact, guiding efficient prompt compression.




\subsection{Type Ablation Analysis}
We categorize tokens into five types, based on the taxonomy proposed by \citealp{natural}: \textbf{Symbol}, \textbf{Signature}, \textbf{Invocation}, \textbf{Identifier}, and \textbf{Structure} (see Appendix \ref{sec:category} for detailed descriptions). 

We constructed Abstract Syntax Trees (ASTs) using JavaParser \citep{javaparser} to identify token types. 
Tokens of specific types were removed from the retrieved code examples, followed by performing RAG with the type-ablated examples to measure their impact on $\mathcal{BLM}$s performance in downstream tasks. The removal priority of a token type $T$ is defined as follows:

\begin{align}
Priority(T) = \frac{\tau_{code/T}}{d_{T}}
\end{align}

\noindent where $\tau_{code/T}$(\%) represents the code compression ratio achieved by discarding tokens of type $T$, 
 $d_{T}$ (\%) denotes the percentage of performance degradation in the evaluation metric caused by the removal of tokens of type $T$.

Token types that yield a higher ratio of $\tau_{code/t}$ and result in minimal degradation $d_{t}$ are assigned higher priority for removal. This approach ensures the compression process effectively reduces token length while preserving the generation quality.

\subsection{Setup of Downstream Coding Tasks with RAG}\label{sec:setup}
To comprehensively assess the impact of different types of tokens, we evaluated them across three datasets and two frozen-parameter $\mathcal{BLM}$s.

\noindent\textbf{Dataset and metric}:  (i) \textbf{\taskone}: The input is a focal method (the method under test) and its partial unit test, while the outputs are assertion statements verifying its correctness. The evaluation metric is the Exact Match Rate, following prior work \citep{UBC}. (ii)\textbf{ \tasktwo}: The input is a buggy method, and the output is a refined version of the method with the bugs fixed. This task is evaluated using CodeBleu \citep{codebleu}, consistent with the CodexGLUE benchmark \citep{codexglue}. (iii) \textbf{\taskthree}: The input consists of a method header (a summary of a function), and the output is a suggested code snippet for the developer based on the header. The task is also evaluated using CodeBleu defined in the original paper \citep{allyouneed}. Table~\ref{tab:task_statistics} presents the statistics of datasets.

\begin{table}[ht]
    \centering
    \resizebox{0.5\textwidth}{!}{%
        \begin{tabular}{lccc}
            \hline
            \textbf{Task}             & \textbf{Knowledge Base (Parsable)} & \textbf{Test} & \textbf{Val} \\ \hline
            \textbf{\taskone}          & 144,112 (70,433)                  & 18,027            & 18,816                  \\
            \textbf{\tasktwo}          & 52,364 (48,903)                   & 6,545             & 6,546                   \\
            \textbf{\taskthree}        & 128,724 (89,014)                  & 10,727            & 5,149                   \\ \hline
        \end{tabular}
    }
    \caption{Dataset statistics of different coding tasks.}
    \label{tab:task_statistics}
\end{table}

In all tasks, we utilize the code RAG prompt template \citep{allyouneed,he2024,UBC} (see Figure \ref{fig:template}), and craft task-specific instructions in a one-shot setting. As listed in Table \ref{tab:task_statistics}, we follow the original split of the dataset into Train, Validation, and Test partitions. The training partition functions as our knowledge base for example retrieval. Note that some code examples that yield parsing errors in JavaParser due to code incompleteness are classified as \textbf{Unparsable}. 
Due to computational resource constraints, we randomly sample 2,000 instances from both the validation and test sets for our experiments. The sampled validation set is used to study example removal priority, while the sampled test set serves to evaluate performance.

.

\noindent\textbf{Base LMs}: The in-context learning capabilities of large language models enable them to utilize query-related documents to produce outputs that better align with the instructions. To investigate whether the impact of different token types is consistent across models, we tested the constructed prompts on two large-scale  $\mathcal{BLM}s$: GPT-3.5-turbo and Gemini-1.0-Pro. We set temperature to 0 to ensure enhanced stability across experiments.

\subsection{Observation}
\begin{figure}
    \centering
    \includegraphics[width=1\linewidth]{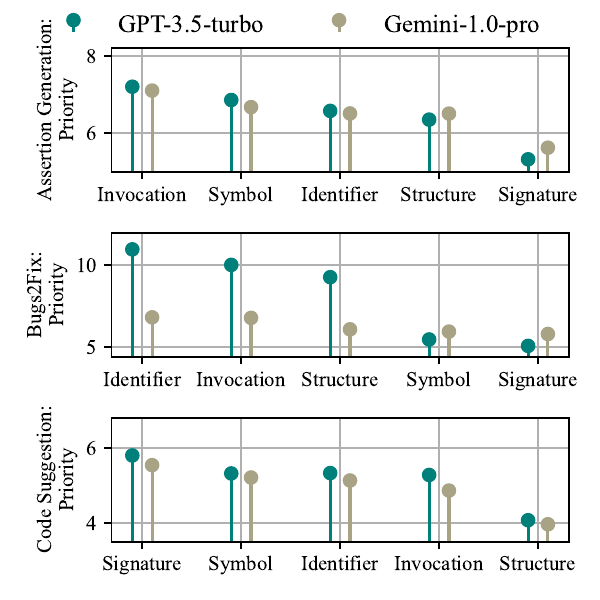}
    \vspace{-0.3in}
    \caption{Removal priority of code token types: e.g., Invocation \textgreater\xspace Symbol in \taskone, and vice versa in \taskthree. Priorities are task-specific yet model-agnostic, applicable to both LMs.}
    \label{fig:rank}
    \vspace{-0.5cm}
\end{figure}

Figure \ref{fig:rank} presents ablation analysis results using a log y-axis to normalize priority scores. The plots reveal type hierarchies of tokens in removal priority, arranged in descending order. This visualization highlights that higher-priority token types should be preferentially removed, providing an intuitive representation of the token-type removal strategy. In addition, the hierarchies are consistent across $\mathcal{BLM}$s but exhibit in-task variations, suggesting the cross-model adaptability of priority-driven code compression.

\section{Methodology}\label{sec:framework}

\begin{figure*}
    \centering
    \includegraphics[width=1\linewidth]{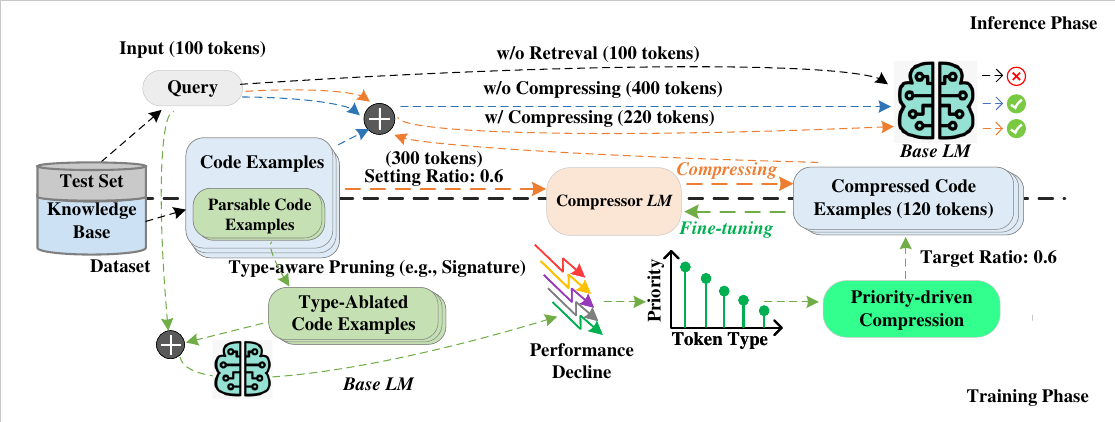}
    \caption{Framework of \ourtool.}
    \label{fig:framework}
    \vspace{-0.3cm}
\end{figure*}

As illustrated in Figure \ref{fig:framework}, \ourtool operates in two phases.
In the training phase, we first derive a type-aware priority ranking for a specific task $\mathcal{T}$ (Sec. \ref{sec:ranking}). Using this ranking, we implement a priority-driven strategy (Algorithm \ref{alg:greedy}): \textit{tokens in higher-priority types are discarded before those in lower-priority types}. This process transforms $(\textbf{\textit{x}}i^{code}, \tau_{code}, \mathcal{T})$ into $\widetilde{\textbf{\textit{x}}}_i^{code}$. We then train $\mathcal{LM_C}$ on the constructed dataset to learn the sequence-to-sequence compression task.

The design of the learning-based $\mathcal{LM_C}$ enhances \textbf{applicability}. While Algorithm \ref{alg:greedy} can directly output compressed code examples, its implementation relies on JavaParser for token labeling and removal, restricting its use to unparsable code. However, as shown in Table \ref{tab:task_statistics},  unparsable code examples are common in coding tasks.

The $\mathcal{LM_C}$ processes code sequences as probabilistic relations \citep{recomp,lingua2} rather than relying strictly on exact syntax, enables our framework to handle both parsable and unparsable code examples while tolerating compile and parse errors \citep{learning}.

In the inference phase, given a query, the $\mathcal{LM_C}$ accepts a specified $\tau_{code}$ and the original retrieved code example to generate compressed code that retains the most critical tokens. These compressed examples are then aggregated into a prompt and passed to the $\mathcal{BLM}$ to generate the final output.



\newcommand{\INPUT}{\item[\textbf{Input:}]}
\newcommand{\OUTPUT}{\item[\textbf{Output:}]}
\begin{algorithm}
\caption{\footnotesize Priority-driven Greedy Algorithm for Dataset Construction} 
\label{alg:greedy}
\begin{algorithmic}[1]
\footnotesize
\INPUT $\textbf{\textit{x}}_i^{code} = \{x_j\}_{j=1}^{L}$, $\tau_{code}$, type priorities of $\mathcal{T}$.
\OUTPUT $\widetilde{\textbf{\textit{x}}}_i^{code}$.
\STATE Initialize a priority queue $\textit{pq}$.
\FOR{each token $x_j \in \textbf{\textit{x}}_i^{code}$}
    \STATE Assign priority to $x_j$ (\textit{Prioritize the drop of high-frequency tokens in prioritized type}).
    \STATE Insert $x_j$ into $\textit{pq}$.
\ENDFOR
\STATE $\textit{removedTokens} \gets \emptyset$.
\STATE $L_{rm} \gets \lfloor \tau_{code} \cdot L \rfloor$.
\STATE $\widetilde{L}_{rm} \gets 0$.
\WHILE{$\widetilde{L}_{rm} < L_{rm}$}
    \STATE $x_j \gets \textit{pq.pop()}$.
    \STATE $\textit{removedTokens} \gets \textit{removedTokens} \cup \{x_j\}$.
    \STATE $\widetilde{L}_{rm} \gets \widetilde{L}_{rm} + 1$.
\ENDWHILE
\STATE $\widetilde{\textbf{\textit{x}}}_i^{code} \gets \textbf{\textit{x}}_i^{code} \setminus \textit{removedTokens}$.
\RETURN $\widetilde{\textbf{\textit{x}}}_i^{code}$.
\end{algorithmic}
\end{algorithm}

\subsection{Code Compression Dataset Construction}\label{sec:trainingset}
The workflow of Algorithm~\ref{alg:greedy} is as follows. Line 1 initializes a priority queue to store tokens alongside their priorities. Lines 2–5 assign priorities to tokens based on their type, with tokens in the same type ranked by Term Frequency (TF) within the example. Frequently occurring tokens are prioritized for removal, as they are more likely to be redundant. Tokens belonging to multiple types are assigned to the category with the lowest removal priority, while out-of-type tokens are removed last, preserving potentially critical tokens. Line 6 initializes an empty set, removedTokens, to track removed tokens. Line 7 calculates the number of tokens to remove as a fraction of the total, determined by the specified $\tau_{code}$. Lines 9–13 iteratively remove the highest-priority tokens from the queue until the required number is removed. Line 14 constructs the modified training sample by excluding tokens in removedTokens from the original sequence. This iterative, priority-driven approach ensures the compressed code retains essential tokens while meeting the specified compression ratio.

Using Algorithm~\ref{alg:greedy}, we constructed a code compression dataset for training compressors (see dataset statistics in Appendix \ref{sec:stat}).


\subsection{Compressor Architecture}


With the code compression dataset, we fine-tune an encoder-decoder model, $\mathcal{LM_C}$, to effectively compress code examples. We adopt CodeT5 \citep{codet5} as our base model and introduce two key modifications to its architecture. First, we extend the input vocabulary with task-indicative tokens such as \textit{<ASSERTION>}, \textit{<BUGS2FIX>}, and \textit{<SUGGESTION>}, which are added at the beginning of the input sequence to explicitly indicate the task context. This design allows our model to be extended to more coding tasks. Additionally, to enable $\mathcal{LM_C}$ to condition on flexible $\tau_{code}$ settings, we introduce special tokens \textit{<Ratio>}, \textit{</Ratio>}, \textit{<Compress>}, and \textit{</Compress>}. These tokens signal the model to generate compressed code snippets tailored to the specified $\tau_{code}$ and task. Moreover, we incorporate a copy mechanism \citep{copy,copy1} into the architecture, allowing the model to directly copy tokens from the input sequence. This modification aligns with the extractive nature of the code compression task, where the outputs are derived entirely from the inputs.

\begin{figure}
    \centering
    \includegraphics[width=1\linewidth]{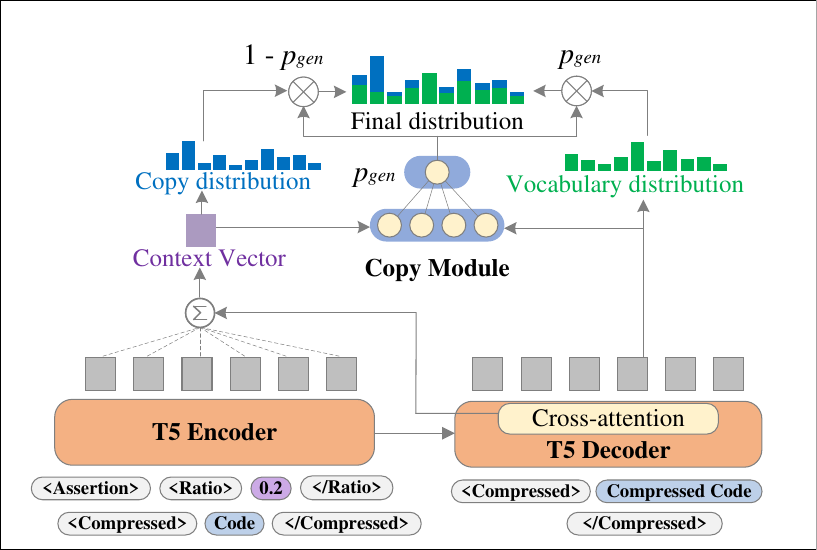}
    \caption{Illustration of copy mechanism on CodeT5.}
    \label{fig:copyT5}
    \vspace{-0.6cm}
\end{figure}

The detailed architecture is shown in Figure \ref{fig:copyT5}. This mechanism is implemented using a copy module that computes the probability of copying each generated token directly from the input, rather than generating it from entire vocabulary. At first, the tokens of the original code sequence $\textbf{\textit{x}}_i^{code} = \{x_j\}_{j=1}^{|\textbf{\textit{x}}_i^{code}|}$  are fed into the encoder, producing a sequence of encoder hidden states $\textbf{h}=\{h_j\}$. In the decoder, the last cross-attention matrix $\mathbf{A} \in \mathbb{R}^{l_{\text{tgt}} \times l_{\text{src}}}$ represents the attention distribution over the source sequence during decoding. Each row $\mathbf{a}^t \in \mathbb{R}^{l_{\text{src}}}$ corresponds to the attention weights assigned to the source sequence at decoding step $t$. Here, $l_{\text{src}}$ and $l_{\text{tgt}}$ denote the maximum input and output lengths, respectively. The attention distribution not only guides the decoder’s focus for each source token, but also allows tokens to be copied from the source sequence by sampling from the attention distribution. Next, the attention distribution generates a weighted sum of the encoder hidden states, known as context vectors $\textbf{h}^*$:
\begin{align}
\textbf{h}_t^*=\sum_i a_i^t \textbf{h}_i, 
\end{align}

The context vector $\textbf{h}_t^*$ represents a fixed size summary of what has been read from the source. Then the context vector is concatenated with the decoder state $\textbf{s}_t$, and passed through a copy module to calculate the generation likelihood $p_{gen} \in [0,1]$ at this step:
\begin{align}
    p_{gen} &= \sigma(\textbf{W}_{gen} \cdot [\textbf{h}_t^*, \textbf{s}_t] +\textbf{b}_{gen})
\end{align}
where $\textbf{W}_{gen}$ and $\textbf{b}_{gen}$ are learnable parameters of the linear copy module. Here, $p_{gen}$ corresponds to the probability of generating tokens from the vocabulary, while $(1-p_{gen})$ denotes the probability of copying tokens from the input.

Next, we calculate the copy distribution by summing the attention weights $a_i^t$ for all positions $i$ where the input token $x_i$ match the target token $y$:
\begin{align}
P_{copy}(y)=\sum_{i: x_i=x} a_i^t
\end{align}

The generation probability is computed through the language model head connected to the decoder’s output, defined as:

\begin{align}
P_{vocab}(y)=Softmax(\textbf{W}_{head} \cdot \textbf{s}_t +\textbf{b}_{head})
\end{align}
where $\textbf{W}_{head}$ and $\textbf{b}_{head}$ denote the weight matrix and bias vector of the head network, respectively.

Finally, the output distribution is computed by interpolating between generation distribution $P_{vocab}$ and copy distribution $P_{\text{copy}}$:

\begin{align}
P(y) = p_{gen}P_{vocab}(y)+(1-p_{gen})P_{copy}(y)
\end{align}

During training, we use the Cross-Entropy Loss to maximize the likelihood of the target sequence. The loss function is defined as:
\begin{equation}
\mathcal{L} = - \sum_{t=1}^{T}y_t \log(\hat{y_t})
\end{equation}

\noindent where $y_t$ is the ground-truth token at step $t$, and $\hat{y_t}$ is the predicted probability of that token. 

We train the model by using the AdamW optimizer with a batch size of 16, a learning rate of 5e-5, and 1,000 warmup steps for 10 epochs.

\section{Experimental Setting}\label{sec:experimentalsetting}



\subsection{Research Questions}
\begin{itemize}
    \setlength\itemsep{-0.07in} 
    \item RQ1: How effective is \ourtool compared to NL-specific prompt compression methods on coding tasks?
    \item RQ2: What is the trade-off between compression ratio and number of shots in \ourtool's performance?
    \item RQ3: How effective is \ourtool in controlling compression ratios?
    \item RQ4: How does \ourtool perform across different $\mathcal{BLM}$s?
    \item RQ5: How does \ourtool perform on unparsable code snippets? 
\end{itemize}

In RQ1, we compare generation quality by the
$\mathcal{BLM}$ using compressed prompts from existing approaches. RQ2 examines the impact of two key hyper-parameters, $\tau_{code}$ and number of shots, analyzing the trade-off between using highly compressed examples versus fewer complete ones within a fixed budget. RQ3 examines the ability of \ourtool on controlling compression ratios. RQ4 evaluates \ourtool's performance across different $\mathcal{BLM}$ to test its generalization. RQ5 explores scenarios with unparsable code, demonstrating the robustness of \ourtool as a learning-based framework.

\subsection{Baselines and Oracle}
We compare our approach against four state-of-the-art prompt compression baselines: LLMLingua \citep{llmlingua}, LongLLMLingua \citep{longllmlingua}, LLMLingua-2 \citep{lingua2}, and RECOMP \citep{recomp}, with detailed descriptions provided in Sec. \ref{sec:related}. For reference, we also evaluate prompts without retrieval or compression. Additionally, we include \emph{Oracle}, where we iteratively remove tokens directly based on their type-aware priority ranking, without using compressor model (as the approach used to construct training dataset in Section~\ref{sec:trainingset}), as a program analysis-based baseline for code examples.

\subsection{Datasets and Metrics}
We evaluated the performance of \ourtool on the same three coding tasks, using the same prompt template and metrics, as presented in Sec. \ref{sec:setup}.

\subsection{Base LMs}
Concrete compression offers the advantage of transferability across various $\mathcal{BLM}$s \cite{recomp,rl}. For RQ1, RQ2, RQ3, and RQ5, we conducted experiments on GPT-3.5-turbo. In RQ4, to evaluate the generalization of \ourtool, we conducted experiments on two additional $\mathcal{BLM}$s: the open-source CodeLlama-13B \citep{codeLlama} and the proprietary LM service Gemini-1.0-pro \citep{gemini}.

\section{Results}\label{sec:results}

\subsection{RQ1: Comparisons with Baselines}\label{sec:rq1}

\begin{table*}[]
\caption{Results on three coding tasks using GPT-3.5-turbo as the $\mathcal{BLM}$. 
To ensure fair comparison with baselines that lack a specified compression rate, we set \ourtool's compression rate to 0.3, keeping it similar to or higher than the baselines. Note that higher metric values indicate better performance, while a higher $\tau$ (\%) reflects a greater proportion of tokens removed from the prompt.}
\centering
\scriptsize
\label{tab:main}
\begin{tabular}{lccccccccc}
\hline
\textbf{}                                             & \multicolumn{3}{c}{\textbf{\taskone}}                                                              & \multicolumn{3}{c}{\textbf{\tasktwo}}                                  & \multicolumn{3}{c}{\textbf{\taskthree}}                  \\ 
\multicolumn{1}{l}{\textbf{Approach}}                  & \textbf{\# tokens} & \textbf{$\tau(\%)$} & {\color[HTML]{242424} \textbf{Exact Match(\%)}} & \textbf{\# tokens} & \textbf{$\tau(\%)$} & \textbf{CodeBleu(\%)} & \textbf{\# tokens} & $\tau(\%)$ & \textbf{CodeBleu(\%)} \\ \hline
\multicolumn{1}{l}{w/o retrieval}                    & 334                & 46.6                         & 23.9                                        & 122                & 66.3                         & 41.7              & 29                 & 82.6                & 14.2               \\ \hline
\multicolumn{7}{l}{\textit{\textbf{Entropy-based}}}                                                                                                                                                                             & \textit{\textbf{}} &                     & \textit{\textbf{}} \\
\multicolumn{1}{l}{LLMLingua}                        & 482                & 22.9                         & 33.8                                        & 286               & 20.9                         & 41.9              & 125                & 25.1                & 21.8               \\
\multicolumn{1}{l}{LongLLMLingua}                    & 474                & 24.2                         & 34.1                                        & 287                & 20.6                         & 42.1              & 126                & 24.1                & 21.2               \\ \hline
\multicolumn{7}{l}{\textit{\textbf{Knowledge Distillation}}}                                                                                                                                                                              & \textit{\textbf{}} &                     & \textit{\textbf{}} \\
\multicolumn{1}{l}{LLMLingua-2}                      & 469                & 25.1                         & 21.2                                        & 282               & 21.9                         & 48.1              & 134                & 19.3                & 21.7               \\
\multicolumn{1}{l}{RECOMP}                           & 465                & 25.6                         & 23.4                                        & 268                & 25.9                         & 45.3              & 132                & 20.9                & 21.0               \\ \hline
\multicolumn{7}{l}{\textit{\textbf{Ours, Setting $\tau_{code}$-0.3, 1-shot}}}                                                                                                                                                     & \textit{\textbf{}} &                     & \textit{\textbf{}} \\
\rowcolor[HTML]{EFEFEF}
\multicolumn{1}{l}{\ourtool w/o Copy} & 447                & 28.5                         & 40.9                                        & 267                & 26.2                         & 56.7              & 131                & 21.7                & 20.5               \\
\rowcolor[HTML]{EFEFEF}
\multicolumn{1}{l}{\ourtool}          & 440                & 29.7                         & 42.1                                        & 262                & 27.4                        & 61.9              & 121               & 27.5                & 23.7               \\ 
\rowcolor[HTML]{F4E5BF}
\multicolumn{1}{l}{Oracle}                           & 454                & 27.4                         & 46.2                                      & 276                & 23.5                         & 66.8              & 120                & 28.1                & 23.8               \\ 
\hline
\multicolumn{1}{l}{w/o Compression}                  & 626                & 0.0                          & 50.5                                        & 362                & 0.0                       & 81.4              & 167                & 0.0                 & 24.7               \\ \hline
\end{tabular}
\vspace{-0.2in}
\end{table*}


Table \ref{tab:main} summarizes the results of different approaches for compressing retrieval-augmented prompts across three coding tasks, evaluating metrics such as token count, $\tau$ (overall compression ratio), and task performance, with the best results reported. The baseline approach without retrieved code examples (w/o retrieval) performs significantly worse than approaches with retrieval, highlighting the importance of RAG in enhancing $\mathcal{BLM}$ performance on coding tasks. \ourtool demonstrates improvements over both entropy-based and distillation-based baselines, improving by 23.4\%, 28.7\%, and 8.7\% over the best baseline for \taskone, \tasktwo, and \taskthree, respectively. 




Comparison with Oracle highlights the learning outcomes derived from the code compression dataset. \ourtool closely approaches Oracle-level performance without requiring JavaParser tools or parsable code snippets, falling only 4.1\% short in Exact Match for \taskone and 4.9\% in CodeBleu for \tasktwo, while achieving nearly identical performance in \taskthree.

The ablation study shows the consistent contributions of the copy mechanism to the enhancement of CodeT5, with a 1.2\% Exact Match increase for \taskone, CodeBleu improvements of 5.2\% for \tasktwo, and 3.2\% for \taskthree. While uncompressed prompts achieve the highest quality metrics, they incur a significant token cost.

\subsection{RQ2: Trade-off between $\tau_{code}$ and Shots}\label{sec:rq2}

\begin{figure}
    \includegraphics[width=1\linewidth]{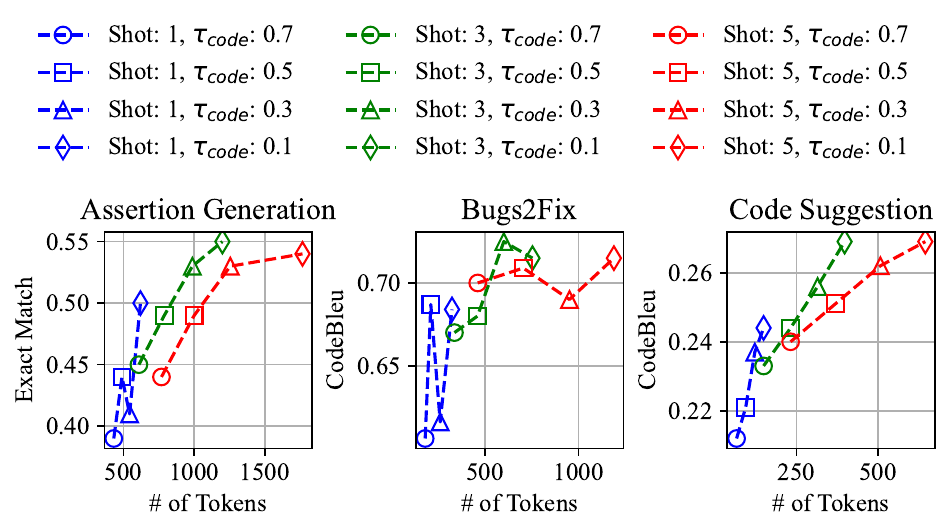}
    \vspace{-0.1in}
    \caption{Trade-off between keeping more tokens in a single example or including more examples.}
    \label{fig:tradeoff}
    \vspace{-0.3cm}
\end{figure}

The objective of prompt compression is to minimize the number of tokens fed to the $\mathcal{BLM}$, while preserving acceptable generation quality. Given a fixed token budget, a trade-off arises between including fewer, less-compressed examples and more highly-compressed ones. Figure \ref{fig:tradeoff} illustrates this balance. \textbf{In general, appending fewer examples, with each example allocated more tokens, achieves better performance than increasing the number of shots while allocating fewer tokens per shot.} For instance, in \taskone, with a token budget of 500, a single example compressed at $\tau_{code}=0.1$ outperforms three examples compressed at $\tau_{code}=0.7$. Additionally, to achieve a fixed performance level, choosing fewer shots with a lower $\tau_{code}$ is more cost-effective, balancing token efficiency and performance.


\subsection{RQ3: Compression Ratio Control}\label{sec:rq2}

\begin{figure}
\subfigure[Assertion Generation ~(b) Bugs2Fix ~ (c) Code Suggestion]
    {\includegraphics[width=1\linewidth]{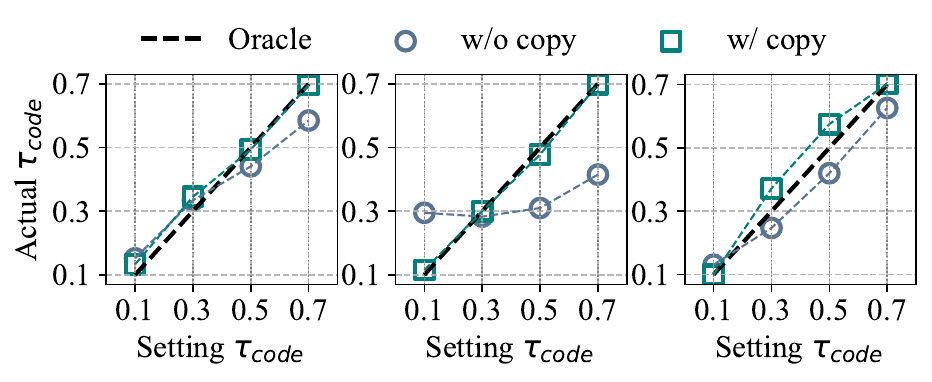}}
    \vspace{-0.2in}
    \caption{Compression ratio control.}\label{fig:ratiocontrol}
    \vspace{-0.2cm}
\end{figure}

Our LM-based compressors utilize an extended vocabulary and accept $\tau_{code}$ as input, enabling adaptive compression of code examples to meet the desired ratio. Figure \ref{fig:ratiocontrol} illustrates the relation between the specified $\tau_{code}$ and the actual achieved values. The dotted line (Oracle) represents the standard outcome, and \ourtool closely aligns with this benchmark.
In contrast, compressors based on the original CodeT5 architecture (w/o the copy) struggle to produce outputs that match the desired ratio. Table~\ref{tab:setting} also provides specific results under varying $\tau_{code}$ configurations, further demonstrating the effectiveness of \ourtool and the critical role of the copy mechanism in achieving accurate compression ratio control.

\subsection{RQ4: Transferability with Different $\mathcal{BLM}$}\label{sec:rq3}

\begin{figure}[tbp]
\subfigure[Assertion Generation ~(b) Bugs2Fix ~ (c) Code Suggestion]
    {\includegraphics[width=1\linewidth]{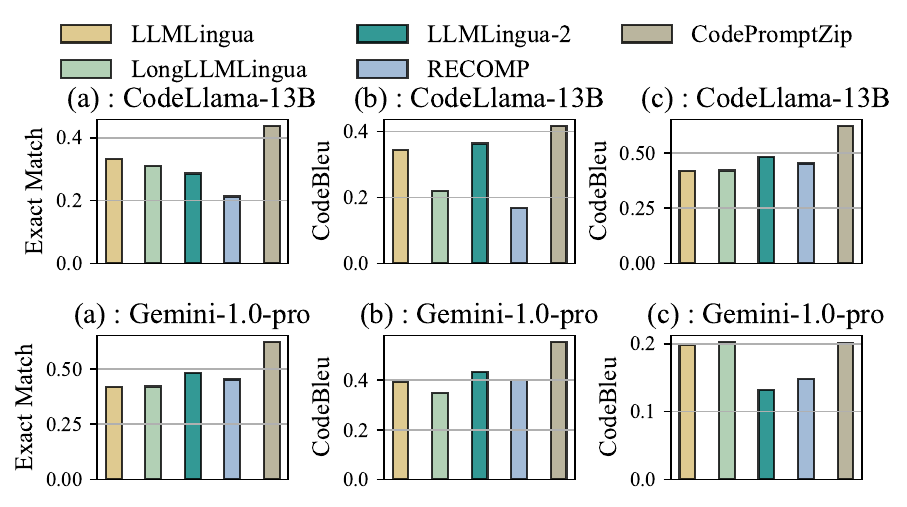}}
    \vspace{-0.2cm}
    \caption{Performance of the proposed \ourtool across different $\mathcal{BLM}$s.}
    \label{fig:llm}
    \vspace{-0.2cm}
\end{figure}

\textbf{\ourtool consistently outperforms baselines across studied base LMs CodeLlama-13B and Gemini-1.0.} Figure \ref{fig:llm} compares the performance among the studied prompt compression techniques across two additional base LMs and all three tasks. In comparison, baseline methods exhibit varying effectiveness, occasionally suffering significant performance drops (e.g., RECOMP on CodeLlama for Bug2fix). The consistent superiority underscores \ourtool’s robustness and effectiveness as a transferable, generalized compression method for code-related tasks.

\subsection{RQ5: Applicability on Unparsable Code}\label{sec:rq4}

Table \ref{tab:unparsable} presents the results of our experiments. Drawing inspiration from \citealp{learning}, we removed a specified percentage of tokens from the end of code examples to render them unparsable and evaluated the effectiveness of our framework. For this experiment, we remove 1\% and 3\% of tokens from the end. When setting $\tau_{code}=0.3$, the exact match rate showed only a slight decrease, from 42.1\% (as shown in Table \ref{tab:main}) for parsable code to 42.0\% and from 42.1\% to 41.7\% when removing 1\% and 3\% of tokens, respectively. In contrast, the Oracle method, which depends on complete code and ASTs, is not applicable (N/A). The results demonstrate the capability of our compressor for real-world scenarios where code completeness cannot always be ensured.

\begin{table}[tbp]
\caption{Results on unparsable code examples.}
\centering
\label{tab:unparsable}
\resizebox{0.5\textwidth}{!}{%
\begin{tabular}{lcccccc}
\hline
\multicolumn{1}{c}{\textbf{}}         & \multicolumn{2}{c}{\textbf{\taskone}}              & \multicolumn{2}{c}{\textbf{\tasktwo}}           & \multicolumn{2}{c}{\textbf{\taskthree}}           \\
\multicolumn{1}{c}{\textbf{Approach}} & \textbf{$\tau(\%)$} & \textbf{Exact Match(\%)} & \textbf{$\tau(\%)$} & \textbf{CodeBleu(\%)} & \textbf{$\tau(\%)$} & \textbf{CodeBleu(\%)} \\ \hline
\multicolumn{7}{l}{\textit{\textbf{Omit 1\% at end, 1-shot Setting code-0.1}}}                                                                             \\
\ourtool w/o Copy                      & 29.1            & 39.7                & 26.4            & 55.4             & 21.9            & 19.4              \\
\ourtool                               & 30.1            & 42.0                & 27.6            & 61.9             & 28.2            & 23.9             \\ \hline
\rowcolor[HTML]{FFCCC9} 
Oracle                                & N/A             & N/A                  & N/A             & N/A               & N/A             & N/A               \\ \hline
\multicolumn{7}{l}{\textit{\textbf{Omit 3\% at end, 1-shot Setting code-0.3}}}                                                                             \\
\ourtool w/o Copy      & 29.5            & 38.4                & 26.5            & 50.8             & 22.0            & 18.7              \\
\ourtool                               & 31.2            & 0.417                & 28.0            & 61.0             & 29.1            & 22.6              \\ \hline
\rowcolor[HTML]{FFCCC9} 
Oracle                                & N/A             & N/A                  & N/A             & N/A               & N/A             & N/A               \\ \hline
\end{tabular}
}
\vspace{-0.2cm}
\end{table}

\section{Conclusion}\label{sec:dis}
This paper presents \ourtool, a framework designed to compress retrieved code examples before incorporating them into prompts. The proposed compressor leverage copy-enhanced LMs and are trained on dedicated datasets. Experimental results demonstrate that \ourtool significantly improves the efficiency of retrieval-augmented LMs while maintaining minimal performance degradation. Note that our framework is not limited to RAG, it could be applied to any prompt that contains code examples.

\section{Limitations}
\textbf{Need for Extra Training for New Coding Tasks:} This study focuses on learning code compression for three method-level coding tasks. Similar to other task-aware compressors (e.g., \citealp{longllmlingua}), the removal priorities in our approach depend on the downstream coding task. If \ourtool{} is to be applied to other coding tasks, such as repository-level tasks \citep{repo}, where token priorities differ significantly, additional training of the compressor is required. However, as shown in Figure \ref{fig:rank}, while priority rankings are task-specific, certain patterns emerge consistently. For example, \textbf{Identifier} tokens exhibit a higher removal priority than \textbf{Structure} tokens across all tasks. We show the out-of-domain capability of our compressor by cross-task experiment in Appendix \ref{sec:out}.

\textbf{Generalizability of Our Findings:}
This study focuses exclusively on Java and method-level tasks. Since programming languages like Python also have program analysis tools, \ourtool{} is applicable to them as well. Future research could extend this work to other tasks and languages. Our experiments utilized GPT-3.5, Gemini, and CodeLlama-13b. We encourage further studies to explore additional base LMs and a broader range of programming languages and coding tasks.

\section{Ethical Considerations}
The implementation of this work is conducted with transparency, providing full disclosure of all technical details, limitations, and potential issues to the relevant stakeholders. The work avoids any false or misleading claims and ensures no data is fabricated or falsified.

In the interest of public benefit, the authors support reasonable and ethical uses of their intellectual contributions. Both the source code and data are released as free and open-source software and are made available in the public domain.

\newpage

\bibliography{main}

\newpage

\appendix
\section{Data Availability}
We have made our replication package available, which contains all the code and datasets available here ~\citep {reproduce}.


\begin{figure}
\subfigure[Assertion Generation ~(b) Bugs2Fix ~ (c) Code Suggestion]
    {\includegraphics[width=1\linewidth]{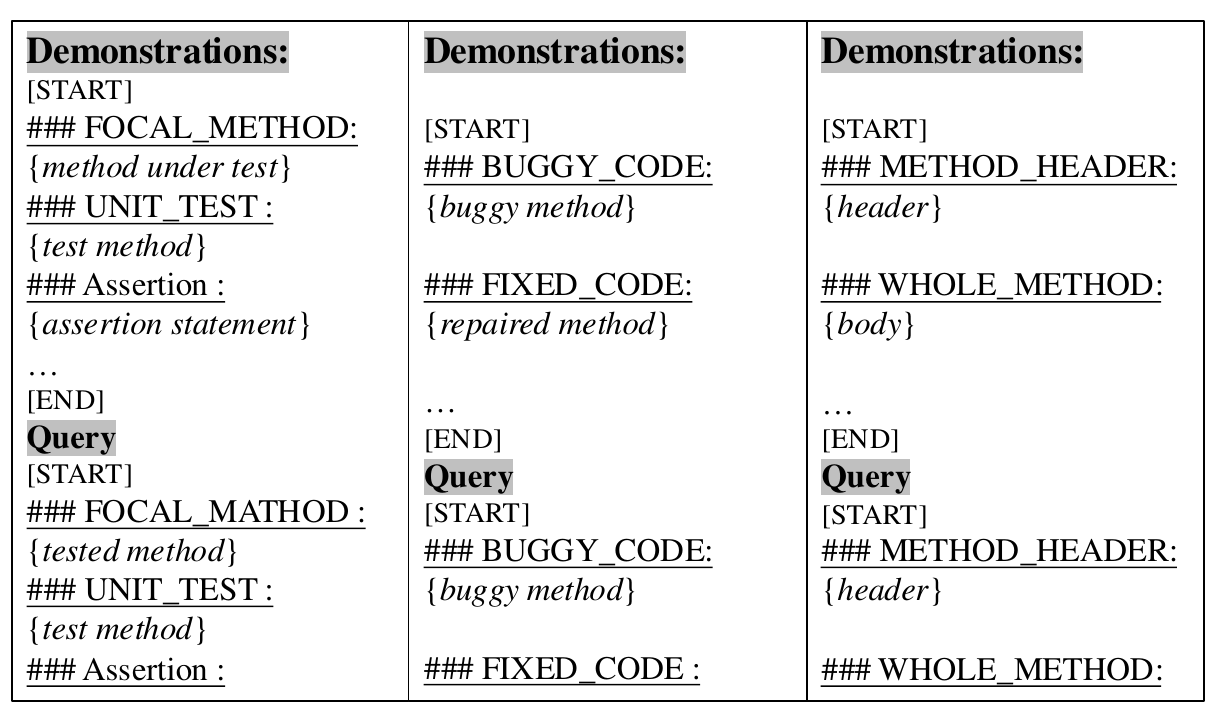}}
    \caption{The illustration of different RAG coding tasks alongside their respective prompt templates.}
    \vspace{-0.3cm}
\end{figure}\label{fig:template}

\section{Token Taxonomy of Code}\label{sec:category}
(i) \textbf{Symbol}: Tokens representing operators, delimiters, and other symbolic elements that define the structure of the code (e.g., {\tt =, \{, ;,}). 
    
\noindent(ii) \textbf{Signature}: Tokens defining the declaration and parameters of methods, critical for understanding the interface and functionality of code components (e.g., {\tt calculate(int x)}).
    
\noindent(iii) \textbf{Invocation}: Tokens related to function or method calls, capturing interactions and dependencies within the code.
    
\noindent(iv) \textbf{Identifier}: Tokens that serve as variable names, class names, or other user-defined labels, are essential for understanding program semantics.

\noindent(v) \textbf{Structure}:  Tokens associated to loops, conditional, and other flow-control statements, which dictate the logical behavior of the program (e.g., {\tt if, for, class}).

\section{Statistics of the Code Compression Dataset for Compressor Training}\label{sec:stat}
The constructed dataset includes original code examples paired with compressed code examples, with $\tau_{code}$ ranging from 0.1 to 0.9. As shown in Table~\ref{tab:stat}, the total number of training samples is nine times the number of parsable code examples in the knowledge base, reflecting the nine distinct $\tau_{code}$ values. The dataset is split into training, validation, and test sets in an 8:1:1 ratio.
\begin{table}[]
\centering
\caption{Statistics of the Code Compression Dataset for Compressor Training.}
\label{tab:stat}
\resizebox{0.5\textwidth}{!}{%
\begin{tabular}{ccc}
\hline
Task       & Total Samples & Split(Training/Test/Validation) \\ \hline
Assertion  & 70433*9       & 80\%/10\%/10\%                  \\
Bugs2Fix   & 48903*9       & 80\%/10\%/10\%                  \\
Suggestion & 89014*9       & 80\%/10\%/10\%                  \\ \hline
\end{tabular}
}
\end{table}

\section{More Results of the Out-of-Domain Capabilities of \ourtool}\label{sec:out}

\begin{table*}[]
\scriptsize
\centering
\caption{Cross-task Results: \textbf{Bold} font indicates the in-task scenario.}
\begin{tabular}{lllcccccc}
\hline
\multicolumn{3}{l}{\textbf{Compressor}} & \multicolumn{2}{c}{(a): \taskone}      & \multicolumn{2}{c}{(b): \tasktwo}   & \multicolumn{2}{c}{(c): \taskthree} \\ \hline
(a)    & (b)    & (c)    & \textbf{$\tau_{code}(\%)$} & \textbf{Exact Match(\%)} & \textbf{$\tau_{code}(\%)$} & \textbf{CodeBleu(\%)} & $\tau_{code}(\%)$      & \textbf{CodeBleu(\%)}   \\ \hline
           \Checkmark       &          &          & \textbf{31.5}       & \textbf{42.1}            & 33.2                & 50.3                  & 19.8            & 13.4                    \\
                  &    \Checkmark        &       & 28.4                & 41.9                     & \textbf{30.0}       & \textbf{61.9}         & 25.1            & 15.9                    \\
                  &          &    \Checkmark      & 34.2                & 34.1                     & 39.9                & 43.6                  & \textbf{32.2}   & \textbf{23.7}           \\ \hline
\end{tabular}
\label{tab:cross}
\end{table*}

To evaluate the out-of-domain effectiveness of our compressors, we performed cross-task experiments using $\mathcal{LM_C}$ trained on individual downstream tasks and tested on the other two out-of-domain tasks. Notably, task-specific special tokens (e.g., <ASSERTION>) were not used in these experiments. Table \ref{tab:cross} summarizes the results of these cross-task evaluations.

For example, in the first row, compressors trained on the Assertion Generation task are applied to compress code examples from Bugs2Fix and Code Suggestion. The Assertion Generation in-task compressor achieves a CodeBleu score of 50.3\% with a $\tau_{code}$ of 33.2\% on Bugs2Fix, compared to its in-task performance of 61.9\% CodeBleu and a $\tau_{code}$ of 30.0\% . While the compressor demonstrates slightly less precision in achieving the desired compression ratio and exhibits degradation in performance, it remains competitive against other baselines, such as the best-performing LLMlingua2 with a CodeBleu score of 48.1\%.

\section{More Results of Impact of the $\tau_{code}$ on the Effectiveness of \ourtool}\label{sec:appendix_ratio}

\begin{table*}[]
\scriptsize
\centering
\caption{Results with varying numbers of code snippets and $\tau_{\text{code}}$ settings of the compressor on the studied tasks. The overall compression ratio $\tau$ is achieved by compressing code snippets with $\tau_{\text{code}}$. }
\label{tab:setting}
\begin{tabular}{lccccccccc}
\hline
\multicolumn{1}{c}{\textbf{}}                       & \multicolumn{3}{c}{\textbf{Assertion Generation}}                          & \multicolumn{3}{c}{\textbf{\tasktwo}}                       & \multicolumn{3}{c}{\textbf{\taskthree}}                     \\
\multicolumn{1}{c}{\textbf{Approach}}               & \textbf{$\tau_{code}(\%)$}   & \textbf{$\tau(\%)$}  & \textbf{Exact Match(\%)} & \textbf{$\tau_{code}(\%)$}   & \textbf{$\tau(\%)$}  & \textbf{CodeBleu(\%)}    & \textbf{$\tau_{code}(\%)$}   & \textbf{$\tau(\%)$}  & \textbf{CodeBleu(\%)}    \\ \hline
\multicolumn{10}{l}{{\color[HTML]{000000} \textit{\textbf{1-shot, Setting $\tau_{code}$-\textcolor{blue}{0.1}}}}}                                                                                                                                                                                            \\
\ourtool w/o Copy                    & \textcolor{blue}{14.2} & 6.2                  & 48.3                 & \textcolor{blue}{29.5} & 25.9                 & 59.0                 & \textcolor{blue}{8.1}  & 5.2                  & 23.4                 \\
\ourtool                             & \textcolor{blue}{13.3} & 7.2                  & 50.1                 & \textcolor{blue}{11.5} & 9.1                  & 68.4                 & \textcolor{blue}{9.9}  & 8.9                  & 24.4                 \\
Oracle                                              & \textcolor{blue}{10.0} & 8.8                  & 49.8                 & \textcolor{blue}{10.0} & 8.1                  & 78.5                 & \textcolor{blue}{10.0} & 8.4                  & 24.5                 \\ \hline
\multicolumn{10}{l}{\textit{\textbf{1-shot, Setting $\tau_{code}$-\textcolor{blue}{0.3}}}}                                                                                                                                                                                                                   \\
\ourtool w/o Copy                    & \textcolor{blue}{33.4} & 28.5                 & 40.9                 & \textcolor{blue}{28.3} & 26.2                 & 56.7                 & \textcolor{blue}{24.7} & 21.7                 & 20.5                 \\
\ourtool                             & \textcolor{blue}{31.5} & 29.7                 & 42.1                 & \textcolor{blue}{30.0}   & 27.4                 & 61.9                 & \textcolor{blue}{32.2} & 27.5                 & 23.7                 \\
Oracle                                              & \textcolor{blue}{30.0} & 27.4                 & 46.2                 & \textcolor{blue}{30.0} & 20.2                 & 66.8                 & \textcolor{blue}{30.0} & 26.1                 & 23.8                 \\ \hline
\multicolumn{10}{l}{\textit{\textbf{1-shot, Setting $\tau_{code}$-\textcolor{blue}{0.5}}}}                                                                                                                                                                                                                   \\
\ourtool w/o Copy                    & \textcolor{blue}{44.1} & 38.3                 & 40.3                 & \textcolor{blue}{31.0} & 21.1                 & 56.5                 & \textcolor{blue}{42.9} & 34.4                 & 23.5                 \\
\ourtool                             & \textcolor{blue}{49.4} & 44.9                 & 45.1                 & \textcolor{blue}{47.8} & 41.4                 & 68.7                 & \textcolor{blue}{57.5} & 42.9                 & 22.1                 \\
Oracle                                              & \textcolor{blue}{50.0} & 45.2                 & 42.1                 & \textcolor{blue}{50.0} & 43.9                 & 67.1                 & \textcolor{blue}{50.0} & 41.9                 & 23.1                 \\ \hline
\textit{\textbf{1-shot}}                            &                              &                      &                      &                              &                      &                      &                              &                      &                      \\
w/o compression                                     & 0.0                          & 0.0                  & 50.5                 & 0.0                          & 0.0                  & 81.4                 & 0.0                          & 0.0                  & 24.7                 \\ \hline
\multicolumn{10}{l}{\textit{\textbf{3-shot, Setting $\tau_{code}$-\textcolor{blue}{0.5}}}}                                                                                                                                                                                                                   \\
\ourtool w/o Copy                    & \textcolor{blue}{44.2} & 39.4                 & 47.2                 & \textcolor{blue}{33.2} & 24.5                 & 67.2                 & \textcolor{blue}{41.2} & 39.1                 & 23.5                 \\
\ourtool                             & \textcolor{blue}{49.3} & 43.8                 & 48.9                 & \textcolor{blue}{50.9} & 42.3                 & 68.0                 & \textcolor{blue}{57.3} & 42.3                 & 24.4                 \\
Oracle                                              & \textcolor{blue}{50.0} & 41.1                 & 52.6                 & \textcolor{blue}{50.0} & 41.9                 & 68.9                 & \textcolor{blue}{50.0} & 42.5                 & 24.6                 \\ \hline
\textit{\textbf{3-shot}}                            & \multicolumn{1}{l}{}         & \multicolumn{1}{l}{} & \multicolumn{1}{l}{} & \multicolumn{1}{l}{}         & \multicolumn{1}{l}{} & \multicolumn{1}{l}{} & \multicolumn{1}{l}{}         & \multicolumn{1}{l}{} & \multicolumn{1}{l}{} \\
w/o compression                                     & 0.0                          & 0.0                  & 55.6                 & 0.0                          & 0.0                  & 85.2                 & 0.0                          & 0.0                  & 24.9                 \\ \hline
\textit{\textbf{5-shot, Setting $\tau_{code}$-\textcolor{blue}{0.5}}} &                              &                      &                      &                              &                      &                      &                              &                      &                      \\
\ourtool w/o Copy                    & \textcolor{blue}{42.9} & 41.3                 & 39.8                 & \textcolor{blue}{30.9} & 28.5                 & 62.4                 & \textcolor{blue}{38.5} & 49.9                 & 24.1                 \\
\ourtool                             & \textcolor{blue}{48.7} & 45.2                 & 49.9                 & \textcolor{blue}{52.1} & 45.1                 & 70.9                 & \textcolor{blue}{49.9} & 41.3                 & 25.1                 \\
Oracle                                              & \textcolor{blue}{50.0} & 43.9                 & 55.2                 & \textcolor{blue}{50.0} & 42.2                 & 74.4                 & \textcolor{blue}{50.0} & 41.6                 & 25.3                 \\ \hline
\textit{\textbf{5-shot}}                            & \multicolumn{1}{l}{}         & \multicolumn{1}{l}{} & \multicolumn{1}{l}{} & \multicolumn{1}{l}{}         & \multicolumn{1}{l}{} & \multicolumn{1}{l}{} & \multicolumn{1}{l}{}         & \multicolumn{1}{l}{} & \multicolumn{1}{l}{} \\
w/o compression                                     & 0.0                          & 0.0                  & 57.8                 & 0.0                          & 0.0                  & 86.4                 & 0.0                          & 0.0                  & 25.6                 \\ \hline
\end{tabular}
\end{table*}

In Table~\ref{tab:setting}, we present results with varying $\tau_{code}$ settings. For compression ratio control, our framework enables the configuration of $\tau_{code}$ as an input to $\mathcal{LM_C}$, allowing it to adaptively compress code examples to match the specified ratio and thereby control the overall $\tau$ of the prompt. However, when using the same configuration and setup with the original CodeT5 structure (w/o copy), the framework struggles to effectively learn the token removal priority. Consequently, the output occasionally deviates from the specified ratio settings, as observed in the \tasktwo 1-shot experiment with $\tau_{code}=0.1$. This underscores the critical role of the copy mechanism in ensuring compliance with ratio settings.

Regarding the quality of generation in the end tasks, the performance varies with different $\tau_{code}$ when a single retrieved example is included. A lower $\tau_{code}$ of 0.1 achieves the highest quality, approximating complete examples. However, lower $\tau_{code}$ values do not always yield better results. For example, in the 1-shot setting on \taskone, $\tau_{code}=0.3$ achieves an exact match rate of 42.1\%, which is lower than the 45.1\% obtained with $\tau_{code}=0.5$ on the same task.

\section{More Results of Impact of the Number of Shots on the Effectiveness of \ourtool}\label{sec:appendix_shot}

We also present results under varying shot settings. The findings indicate that increasing the number of compressed examples improves performance, consistent with prior observations for complete examples \citep{he2024}. Notably, this improvement extends to compressed examples. For instance, in \taskthree, increasing the number of $\tau_{code}=0.5$ compressed examples from 1 to 5 raises the CodeBleu score from 22.1 to 25.1, highlighting the advantage of incorporating multiple compressed examples into the prompt.

\section{Case Study}\label{sec:appendix_casestudy}

We present cases across multiple coding datasets, comparing compressed and original code examples. For instance, as demonstrated in Figures 8, 9, and 10, \ourtool prioritizes discarding \textbf{Invocation} tokens first, followed by \textbf{Symbol} tokens.
\begin{figure}[!h]
\begin{tcolorbox}
\begin{lstlisting}[language=Java,frame=single,framerule=0pt]
### FOCAL_METHOD 
getProduction(java.lang.String) { 
 return productionsByName.get(name); }  
### UNIT_TEST  
testJustifications() { 
 runTest("testJustifications", 2); org.jsoar.kernel.Production j = agent.getProductions() .getProduction("justification-1"); "<AssertPlaceHolder>"; 
}    
\end{lstlisting}
\end{tcolorbox}
\caption{Original Code Examples of Assertion Generation (63 tokens)}
\label{fig:code-example}
\end{figure}

\begin{figure}[!h]
\begin{tcolorbox}
\begin{lstlisting}[language=Java,frame=single,framerule=0pt]
### FOCAL_METHOD 
getProduction(java.lang.String) { 
 return productionsByName; }  
### UNIT_TEST  
testJustifications() { 
 ; 
 org.jsoar.kernel.Production j = agent.getProductions() .getProduction("justification-1"); "<AssertPlaceHolder>"; 
}    
\end{lstlisting}
\end{tcolorbox}
\caption{Compressed Code Examples of Assertion Generation (55 tokens, $\tau_{code}$: 0.1)}
\label{fig:code-example}
\end{figure}

\begin{figure}[!h]
\begin{tcolorbox}
\begin{lstlisting}[language=Java,frame=single,framerule=0pt]
### FOCAL_METHOD 
getProduction(java.lang.String)  
 return productionsByName;     
### UNIT_TEST  
testJustifications()  
 ; 
 org.jsoar.kernel.Production j = agent;
  "<AssertPlaceHolder>"; 
\end{lstlisting}
\end{tcolorbox}
\caption{Compressed Code Examples of Assertion Generation (39 tokens, $\tau_{code}$: 0.4)}
\label{fig:code-example}
\end{figure}

\begin{figure}[!h]
\begin{tcolorbox}
\begin{lstlisting}[language=Java,frame=single,framerule=0pt]
### BUGGY_CODE 
public static TYPE_1 init(java.lang.String name, java.util.Date date) {
   TYPE_1 VAR_1 = new TYPE_1();
   VAR_1.METHOD_1(name);
   java.util.Calendar VAR_2 = java.util.Calendar.getInstance();
   VAR_2.METHOD_2(date);
   VAR_1.METHOD_3(VAR_2);
   return VAR_1;
}
### FIXED_CODE   
public static TYPE_1 init(java.lang.String name, java.util.Date date) {
   TYPE_1 VAR_1 = new TYPE_1();
   VAR_1.METHOD_1(name);
   java.util.Calendar VAR_2 = null;
   if (date != null) {
       VAR_2 = java.util.Calendar.getInstance();
       VAR_2.METHOD_2(date);
   } 
   VAR_1.METHOD_3(VAR_2);
   return VAR_1;
}
\end{lstlisting}
\end{tcolorbox}
\caption{Original Code Examples of Bugs2Fix (195 tokens)}
\label{fig:code-example}
\end{figure}

\begin{figure}[!h]
\begin{tcolorbox}
\begin{lstlisting}[language=Java,frame=single,framerule=0pt]
### BUGGY_CODE 
public static TYPE_1 init(java.lang.String name, java.util.Date date) {
    = new TYPE_1();
   ;
   java.util.Calendar = java.util.Calendar;
   .METHOD_2(date);
   .METHOD_3(VAR_2);
   return ;
}
### FIXED_CODE   
public static TYPE_1 init(java.lang.String name, java.util.Date date) {
    = new TYPE_1();
   ;
   java.util.Calendar = null;
   if (date != null) {
        = java.util.Calendar;
       .METHOD_2(date);
   } 
   .METHOD_3(VAR_2);
   return ;
}
\end{lstlisting}
\end{tcolorbox}
\caption{Compressed Code Examples of Bugs2Fix (136 tokens, $\tau_{code}$: 0.3)}
\label{fig:code-example}
\end{figure}

\begin{figure}[!h]
\begin{tcolorbox}
\begin{lstlisting}[language=Java,frame=single,framerule=0pt]
### METHOD_HEADER 
protected final void fastPathEmit ( U value , boolean delayError , Disposable dispose )
### WHOLE_METHOD  
protected final void fastPathEmit(U value, boolean delayError, Disposable dispose) {
   final Observer<? super V> s = actual;
   final SimplePlainQueue<U> q = queue;
   if (wip.get() == 0 && wip.compareAndSet(0, 1)) {
       accept(s, value);
       if (leave(-1) == 0) {
           return;
       }
   } else {
       q.offer(value);
       if (!enter()) {
           return;
       }
   }
   QueueDrainHelper.drainLoop(q, s, delayError, dispose, this);
}
\end{lstlisting}
\end{tcolorbox}
\caption{Original Code Examples  of \taskthree (157 tokens, $\tau_{code}$: 0.3)}
\label{fig:code-example}
\end{figure}

\begin{figure}[!h]
\begin{tcolorbox}
Original Code Examples (121 tokens, $\tau_{code}$: 0.3)
\begin{lstlisting}[language=Java,frame=single,framerule=0pt]
### METHOD_HEADER 
protected final void fastPathEmit ( U value , boolean delayError , Disposable dispose )
### WHOLE_METHOD  
   final Observer<? super V> = 
   final SimplePlainQueue<U> = 
   if (wip.get() == 0 && wip.compareAndSet(0, 1)) 
       ;
       if (leave(-1) == 0) 
           return;    
    else 
       .offer(value);
       if (!enter()) 
           return;
   .drainLoop(q, s, delayError, dispose, this);
\end{lstlisting}
\end{tcolorbox}
\caption{Compressed Code Examples  of \taskthree (121 tokens, $\tau_{code}$: 0.3)}
\end{figure}

\end{document}